\documentclass{vgtc}                          

\usepackage{times}

\usepackage{tabu}                      
\usepackage{booktabs}                  
\usepackage{multirow}

\usepackage{mathptmx}                  
\usepackage{amsmath}
\usepackage{amssymb}

\usepackage{enumitem}

\usepackage{subcaption}
\usepackage{fontawesome6}
\usepackage{tikz}

\usepackage[hyphens]{url}

\newcolumntype{L}[1]{>{\raggedright\arraybackslash}p{#1}}
\newcolumntype{C}[1]{>{\centering\arraybackslash}p{#1}}
\newcolumntype{R}[1]{>{\raggedleft\arraybackslash}p{#1}}

\usetikzlibrary{arrows.meta,chains,shapes.symbols,positioning}
\newif\ifurbanprocessA

\usepackage{comment}

\vgtcinsertpkg

\title{Image Generation Techniques for Urban Planning}

\author{Katharina Roth\thanks{e-mail: kroth@rptu.de}\\ 
        \scriptsize University of Kaiserslautern-Landau 
\and Eva Hagen\thanks{e-mail: eva.hagen@rptu.de}\\ 
		\scriptsize University of Kaiserslautern-Landau
\and Alexander Bartscher\thanks{e-mail: alexander.bartscher@rptu.de}\\ 
		\scriptsize University of Kaiserslautern-Landau
\and Inga Scheler\thanks{e-mail: scheler@rptu.de}\\ 
		\scriptsize University of Kaiserslautern-Landau
\and Nicolas R. Gauger\thanks{e-mail: nicolas.gauger@rptu.de}\\ 
		\scriptsize University of Kaiserslautern-Landau}

\abstract{
	In the context of urban planning, architects are normally instructed with creating presentation images that visualize proposed buildings within their urban context. This work aims to develop a GenAI model for automatically generating architectural presentation images in urban scenes, with emphasis on model optimization. To achieve this, we developed  Mask-based Weighted Conditional Flow Matching~(MWCFM), which extends Flow Matching by introducing contextual masks for precise feature focusing. This enables targeted training on critical spatial elements relevant to urban planning. Our trained model learns from urban street-view data while adhering to specific style-guidelines, which are integrated into training through the loss function. Furthermore, the model's performance is evaluated using application related metrics, derived from presentation image style guidelines.
}


\begin{document}

\maketitle


\section{Introduction}\label{ch:intro}

The interest in using \textit{Generative Artificial Intelligence} (GenAI) models for generating new content has increased dramatically over the past decade~\cite{gozalo2023survey}. Prominent examples include \textit{Large Language Models (LLMs)} such as OpenAI's \textit{GPT}, utilized for text generation, as well as image generation models like \textit{Midjourney} or \textit{DALL-E}, based on text-to-image generation technologies~\cite{kenthapadi2023generative}. 

Beyond artificial intelligence research, the attention on this topic is growing across diverse domains~\cite{kenthapadi2023generative, gozalo2023survey}, including medicine~\cite{teo2025generative}, geology~\cite{ghyselincks2025synthetic}, and urban planning~\cite{liu2024day}, whereby the model requirements often depend significantly on the application area. This work focuses on GenAI models for application task in the urban planning domain, more precisely, on an architectural design task in urban areas.

\begin{figure}[!h]
	\centering
	\includegraphics[width=\linewidth]{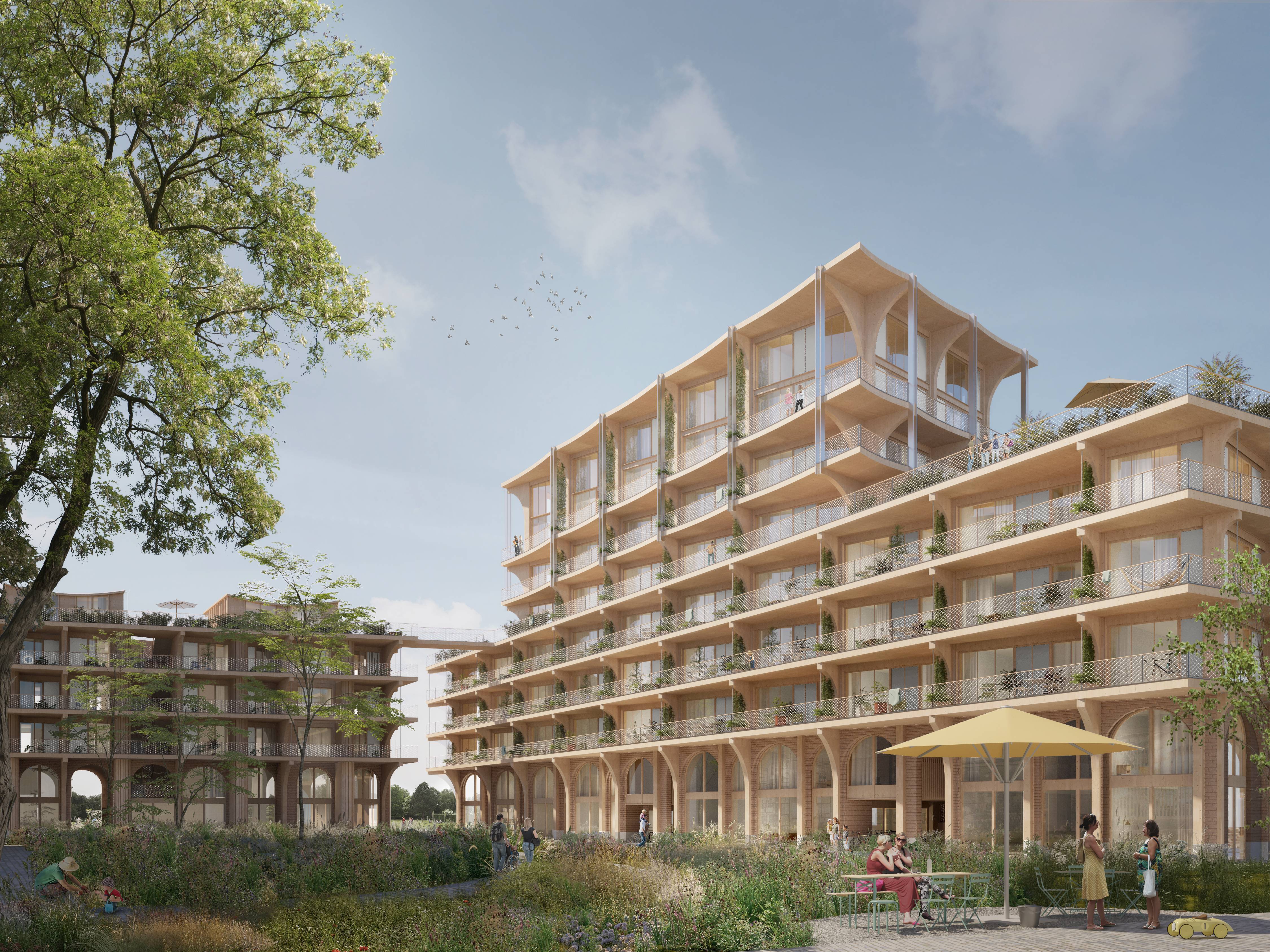}
	\caption[Example design draft of an presentation image]{Example design draft of an presentation image provided by Bartscher~\cite{Bartscher2025}. Here, all style-guidelines [C1]-[C7] (see Section~\ref{chs:styleguidelines}) are fulfilled.´}
	\label{fig:optimum}
\end{figure}

During \textit{design phase} within the so-called \textit{urban planning process}, architects can be instructed to design a building. When a project is commissioned, architects develop a presentation draft that visualizes the designed building within its intended urban context, rendering the proposed architectural structures including the surrounding cityscape. Regarding such \textit{presentation images} several style-guidelines exists. Figure~\ref{fig:optimum} exemplifies such guideline-compliant imagery. Further details on the urban planning process and style-guidelines are discussed in Section~\ref{ch:urban}. Creating such presentation images is very time-consuming, motivating this work to support architects and urban planners through GenAI.
\vspace{0.5cm}

This results in the following three issues:
\begin{enumerate}
	\item How to automatically create urban presentation images or parts of it using GenAI models?
	
	\item How to guarantee compliance of urban style guidelines during training?
	
	\item How to develop and optimize GenAI models which are capable of producing these images?
\end{enumerate}

So, the overall objective is to automatically generate images of buildings within urban scenes using GenAI techniques that most comprehensively meet the style-guidelines. However, the primary focus of this work is not on creating such presentation images, but rather on highlighting the process of development and optimization of a GenAI model capable of producing these images regarding the application-specific style-guidelines via a prototype model.

A significant semantic challenge in training is that users cannot control which features or image regions are more relevant to the application than others. Addressing this problem, we introduce the developed \textit{Mask-based Conditional Weighted Flow Matching~(MWCFM)} approach, an advanced generative model, an adaption of the \textit{ Weighted Conditional Flow Matching~(WCFM)} approach, developed by Calvo-Ordoñez et al.~\cite{calvo2025weighted}, by enabling focused training on application-specific features. By integrating innovative weighting techniques, MWCFM allows expert knowledge to be incorporated into the training process, achieving precise feature focusing on critical contextual elements. We demonstrate its effectiveness through urban planning image generation, where the model prioritizes contextual elements in spatial visualization. Model's Performance is evaluated using metrics derived from the presentation image style-guidelines, assessing image generation quality, contextual accuracy, and visual representation precision in urban planning contexts.

This paper is structured as follows: After presenting related Work and basic foundations about \textit{Generative Artificial Intelligence~(GenAI)} in Section~\ref{ch:related}, basic principles and style-guidelines of urban planning, especially regarding image generation of presentation images, are presented in Section~\ref{ch:urban}. Then, in Section~\ref{ch:image}, our prototype model training, based on our MWCFM approach, is presented. Afterwards, the evaluation techniques and metrics, derived from urban style-guidelines, are explained in Section~\ref{ch:eval}. Then, the evaluation and training results are presented in Section~\ref{ch:results} and discussed in Section~\ref{ch:discussion}. Finally, we conclude our work and highlight future work in Section~\ref{ch:con}. 

\section{Related Work}\label{ch:related}

In this section the basic foundations and related work on GenAI, developed within the urban planning domain, are illustrated.

\subsection{Generative Artificial Intelligence (GenAI)}

\subsubsection{Definition}

\textit{Generative Artificial Intelligence}~(GenAI) is a machine learning approach in which models, so-called \textit{GenAI models}, are designed to learn the underlying structure of a dataset and automatically generate new, plausible data points resembling the original dataset~\cite{bordas2024generative}. If those data points are images we speak of image generation. 

Mathematically, a generative model is defined by a joint probability distribution $ p(x) $ over datapoints $ x $ within a potentially high-dimensional space $ X $, characterizing the underlying data structure~\cite{alma992022090961107476, doersch2016tutorial}. When applied to images, this is called image generation~\cite{alma992022090961107476, doersch2016tutorial}. Furthermore, a \textit{conditional generative model} $ p(x|c) $ is a specific type of a generative model which includes conditions on inputs or covariances, enabling more targeted generation~\cite{alma992022090961107476}. 

\subsubsection{Generative Model Types}
 
There exists various types of generative models, whereas the definition of GenAI applies across all of them, though interpretation varies by model. According to Murphy~\cite{alma992022090961107476} six of these model types are frequently used and can be seen as the \textit{main types}. We added \textit{Flow Matching} into this list, due to its rising popularity in the last few years~\cite{lipman2022flow, lipman2024flow, calvo2025weighted, gat2024discrete}:
\begin{itemize}
	\item \textbf{Energy-Based Model (EBM)}~\cite{lecun2006tutorial, alma992022090961107476, zhai2016deep}:\\
	Dependencies between variables are expressed through an energy function.

	\item \textbf{Generative Adversarial Network (GAN)}~\cite{arora2017theoretical, you2022application}: \\
	An generator network is trained to generate synthetic data and the discriminator network is trained to distinguish between real or fake data.
	
	\item \textbf{Variational Autoencoder~(VA)}~\cite{connor2021variational, girin2020dynamical}:\\
	Transforms the input data into a probabilistic latent vector representation and decodes them in a second step to learn the distribution of the data.
	
	\item \textbf{Autoregressive Model~(ARM)}~\cite{alma992022090961107476, bond2021deep, hoogeboom2021autoregressive}: \\
	Predict each element of a sequence conditioned on previous elements.
	
	\item \textbf{Continuous Flow Network~(CFN)}~\cite{alma992022090961107476, ho2019flow++, zhen2021flow}:\\
	 Learn an invertible mapping between a simple latent distribution and the data distribution.
	
	\item \textbf{Diffusion Model~(DM)}~\cite{alma992022090961107476, croitoru2023diffusion, fishman2023diffusion}:\\
	Gradually add noise to data, turning it into a simple distribution and then learn to denoise it again step‑by‑step, reconstructing data from pure noise.
	
	\item \textbf{Flow Matching Models (FM)}~\cite{lipman2022flow}:\\
	 Learn the time-dependent velocity field that transforms a known source distribution (e.g. Gaussian) into a target distribution by integrating an Ordinary Differential Equation~(ODE) over time.
\end{itemize}

Regarding image generation, diffusion models offer stable training, high-quality images, and good scalability~\cite{croitoru2023diffusion}, but require substantial time- and computational resources~\cite{wang2025generative}. Flow Matching offers the advantages of diffusion models while requiring less training time and also leverages the benefits of CFNs, which is why, despite its implementation complexity, it is our chosen approach~\cite{research_Report_FM, simple_science_2025}.

\subsubsection{Flow Matching}

The idea begind FM models is to construct a time-dependent velocity field that transforms source distributions into target distributions via ODE integration, learning the velocity field directly along probability paths rather than through the iterative noise perturbation characteristic of diffusion models~\cite{fishman2023diffusion, lipman2022flow}. However, FM's loss function is computationally intractable in practice, because it requires integration over the entire unknown data distribution to determine the velocity field and corresponding density. This constraint cannot be satisfied when only individual data points, rather than the exact density function, are available~\cite{lipman2022flow}. 

\textit{Conditional Flow Matching}~(CFM)~\cite{lipman2022flow} resolves this by decomposing the global intractable problem into tractable sub-problems conditioned on single data samples and corresponding noise samples. Building upon this approach, recent work provided by Calvo-Ordóñez et al.~\cite{calvo2025weighted} has incorporated weighting mechanisms that prioritize samples based on their information gain, enabling the model to learn the conditional flow more effectively by controlling each sample's contribution to the loss.

However, when considering our urban planning image generation task, there is a key limitation. During training, users cannot control which features or image regions are more relevant to the application than others. This represents a semantic challenge that requires expert knowledge to incorporate into the training process. Addressing this problem motivated the development of our MWCFM which is an extended version of the approach from Calvo-Ordoñez et al.~\cite{calvo2025weighted}.

\subsection{GenAI in Urban Planning}

Due to the multifaceted nature of urban planning, an equally diverse range of applications in this field deals with the generation of new content~\cite{gozalo2023survey}. Because urban planners tackle complex, interconnected challenges across various domains, the potential uses of GenAI are correspondingly broad and adaptable~\cite{gozalo2023survey, MARASINGHE2024105047}. The academic research community has recognized this potential~\cite{kenthapadi2023generative, liu2024day}. Therefore, numerous papers are published in recent years that aim to support urban planners in their daily work. Examples of such paper are presented in this section.

\subsubsection{Urban Planning Generation Tasks}

There exist many studies in the area of urban planning across various applications. For example Liu et al.~\cite{liu2024day} support urban planners in analyzing potential security risks within a city. They developed a GenAI model that can transform street view images from day to night. Furthermore,  Wang et al.~\cite{wang2023automated} developed a model for automated generation of land-use configurations. The provided model is able to reconfigurate urban structures by generating complex urban planning patterns while simultaneously quantitatively evaluating their quality and functionality. Zheng et al.~\cite{zheng2021generative} developed a GenAI tool for architectural design in urban environments that learns design patterns to produce precise 3D vector geometries.

\subsubsection{Image Generation for Urban Planning Tasks}

When looking into image generation tasks, several types of GenAI models can be used. The most popular model one are diffusion models. Due to their high potential, various applications for urban planners were developed in the last few years. 

For example Wang et al~\cite{wang2025generative} used diffusion models in combination with a controlNet to generate high-resolution satellite images based on descriptions of landscape, infrastructure and natural environment. He et al.~\cite{he2025generative} developed a GenAI framework also using diffusion models for generating  various designs. This approach is based on human expert knowledge included via textual prompts and image-based constraints. Cui et al.~\cite{cui2024learning} explore an approach to urban design using a conditional latent diffusion model for generating urban design images. Here, diffusion models were used to automatically create photo-realistic and contextually appropriate synthesized urban design visualizations based on a conditional control network.

For our best of knowledge, application based paper regarding urban planning image generation using FM are rather rare.

\section{Urban Planning} \label{ch:urban}

This section explores urban planning as the application field and examines the style-guidelines that govern GenAI-based image generation of buildings in urban contexts, which form the primary focus of this work.

\urbanprocessAfalse 
\ifurbanprocessA
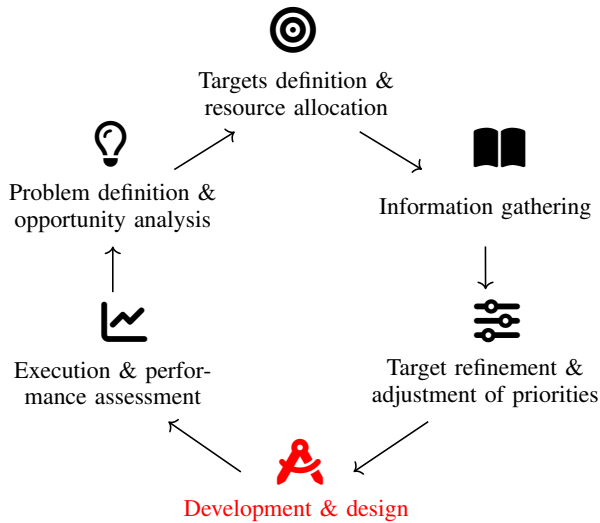
\begin{figure*}[h]
	\centering
	\begin{tikzpicture}
		\tikzset{
			mynode/.style={
				rectangle,  
				line width=.5pt, 
				minimum width=3cm, 
				minimum height=1cm,
				text width=3.2cm,
				align=center
			},
			mini/.style={
				rectangle, 
				line width=1.5pt, 
				minimum width=0.3cm, 
				minimum height=0.3cm,
				text width=0.3cm,
				align=center
			},
			myarrow/.style={
				->, line width=0.6pt
			}
		}
		\node[mynode] (step1) {Problem definition \& opportunity analysis }; 
		\node[mini,above=-.1cm of step1] (id1){\huge {\faLightbulb[regular]}};
		
		\node[mynode, above right=-0.1 and .3cm of step1] (step2) {Targets definition \& resource allocation}; 
		\node[mini, , above=-.1cm of step2] (id2){\huge{\faBullseye}};
		
		\node[mynode, below right=.1cm and 0.3cm of step2] (step3) {Information gathering}; 
		\node[mini, above=-.1cm of step3] (id3){\huge{\faBookOpen}};
		
		\node[mynode, below=1.3cm of step3] (step4) {Target refinement \& adjustment of priorities}; 
		\node[mini, above=-.1cm of step4] (id4){\huge{\faSliders}};
		
		\node[mynode, text=red, below=3cm of step2] (step5) {Development \& design}; 
		\node[mini, text=red, above=-.1cm of step5] (id5){\huge{\faCompassDrafting}};
		
		\node[mynode, below=1.3cm of step1] (step6) {Execution \& performance assessment}; 
		\node[mini,above=-.2cm of step6] (id6){\huge{\faChartLine}};
		
		\draw[myarrow] (step1) -- (step2);
		\draw[myarrow] (step2) -- (step3);
		\draw[myarrow] (step3) -- (id4);
		\draw[myarrow] (step4) -- (step5);
		\draw[myarrow] (step5) -- (step6);
		\draw[myarrow] (id6) -- (step1);
	\end{tikzpicture}
	\caption[Urban planning process cycle]{Urban planning process cycle regarding Marasinghe et al.~\cite{MARASINGHE2024105047}, representing the basic workflow of urban planners. GenAI can be used across all steps for support. Image generation of \textit{presentation images} belongs to design step~(\faCompassDrafting), highlighted in red.}
	\label{fig:up_process}
\end{figure*}
\else

\begin{figure}[!b]
	\centering
	\begin{tikzpicture}
		\tikzset{
			mynode/.style={
				rectangle,  
				line width=.5pt, 
				minimum width=3cm, 
				minimum height=1cm,
				text width=3.2cm,
				align=center
			},
			mini/.style={
				rectangle, 
				line width=1.5pt, 
				minimum width=0.3cm, 
				minimum height=0.3cm,
				text width=0.3cm,
				align=center
			},
			myarrow/.style={
				->, line width=0.6pt
			}
		}
		\node[mynode] (step1) {Problem definition \& opportunity analysis }; 
		\node[mini,above right=-.1cm and -2.1cm of step1] (id1){\huge {\faLightbulb[regular]}};
		
		\node[mynode, above right=.5cm and -1cm of step1] (step2) {Targets definition \& resource allocation}; 
		\node[mini, , above left =-.1cm and -1.8cm of step2] (id2){\huge{\faBullseye}};
		
		\node[mynode, right=1.5cm of step1] (step3) {Information gathering}; 
		\node[mini, above=-.1cm of step3] (id3){\huge{\faBookOpen}};
		
		\node[mynode, below =1.3cm of step3] (step4) {Target refinement \& adjustment of priorities}; 
		\node[mini, above left=-.1cm and -2cm of step4] (id4){\huge{\faSliders}};
		
		\node[mynode, text=red, below=4.5cm of step2] (step5) {Development \& design}; 
		\node[mini, text=red, above right=-.3cm and -2.1cm of step5] (id5){\huge{\faCompassDrafting}};
		
		\node[mynode, below=1.3cm of step1] (step6) {Execution \& performance assessment}; 
		\node[mini,above=-.1cm of step6] (id6){\huge{\faChartLine}};
		
		\draw[myarrow] (step1) -- (step2);
		\draw[myarrow] (step2) -- (step3);
		\draw[myarrow] (step3) -- (id4);
		\draw[myarrow] (step4) -- (step5);
		\draw[myarrow] (step5) -- (step6);
		\draw[myarrow] (id6) -- (step1);
	\end{tikzpicture}
	\caption[Urban planning process cycle]{Urban planning process cycle regarding Marasinghe et al.~\cite{MARASINGHE2024105047}, representing the basic workflow of urban planners. GenAI can be used across all steps for support. Image generation of \textit{presentation images} belongs to design step~(\faCompassDrafting), highlighted in red.}
	\label{fig:up_process}
\end{figure}
\fi

\subsection{Definition}

\begin{figure*}[!h]
	\centering
	\includegraphics[width=0.8\linewidth]{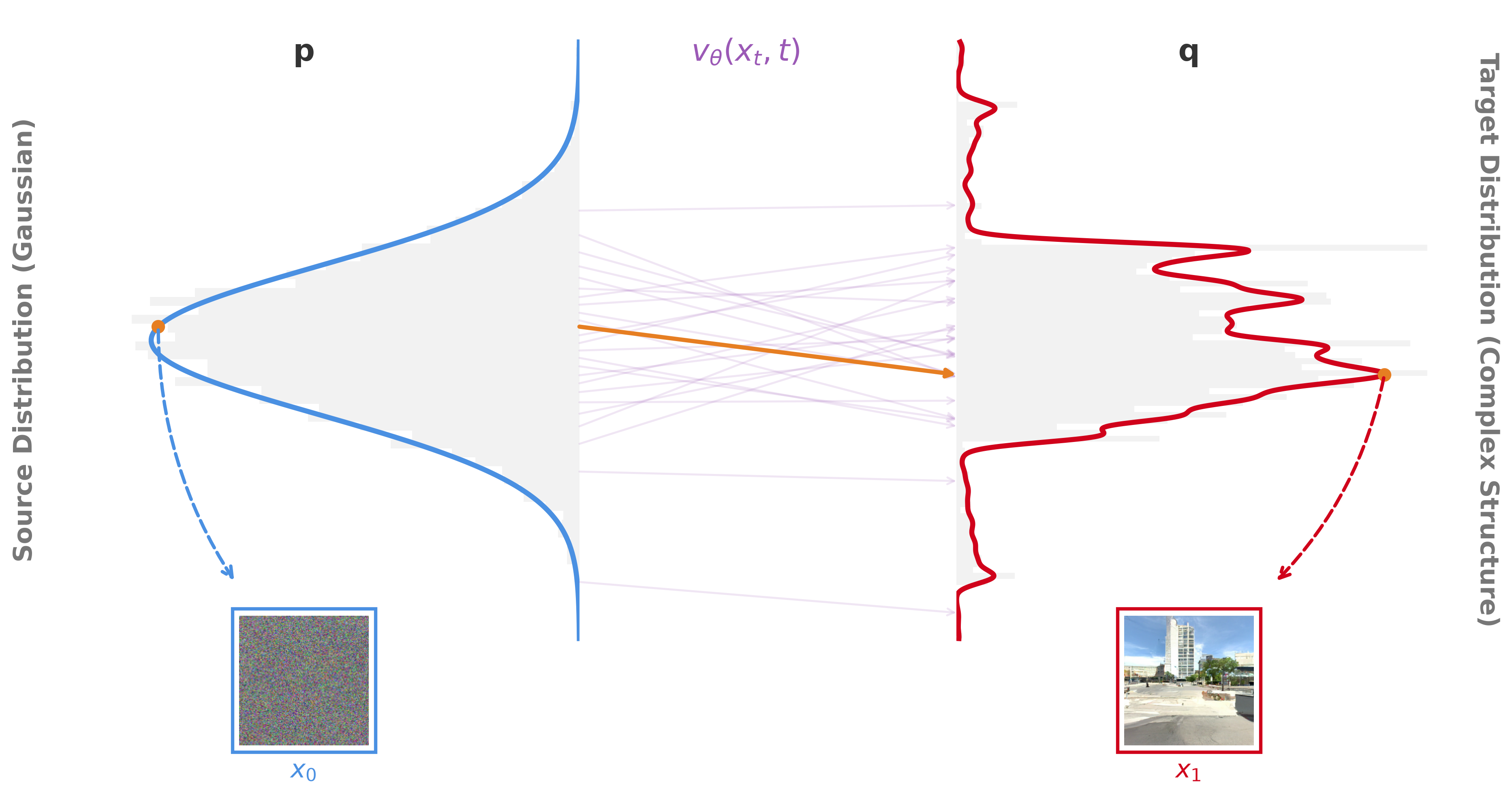}
	\caption{Illustration of Conditional Flow Matching process as described by Lipman et al~\cite{lipman2022flow}. Here, the continuous transport from a source distribution~$p$~(blue), represented by noisy input samples~$x_{0}$, to a target distribution~$q$~(red), represented by target samples~$x_{1}$. The learned flow, depicted by arrows, indicating the trajectory of the samples under the conditional vector field over time.}
	\label{fig:fm}
\end{figure*}

According to Bibri et al.~\cite{bibri2017smart} \textit{urban planning} can be defined as 'a complex discipline aimed at designing and developing urban spaces in a comprehensive and forward-looking manner. It combines six main perspectives from social, cultural, political, economic, physical, and ecological realms'. This means that urban planners try to improve urban spaces within a city regarding these six perspectives simultaneously. Therefore, the main task of an urban planner is the management and development of land use, urban environment, and infrastructure to maximize the benefit of all perspectives. In practice, this is a challenging task that involves requiring a systematic approach, the so-called \textit{urban planning processes cycle}~\cite{MARASINGHE2024105047}.

Marasinghe et al.~\cite{MARASINGHE2024105047} provide a basic version of such an urban planning process cycle, consisting of six steps, shown in Figure~\ref{fig:up_process}. In the beginning, urban planners identify issues and analyze opportunities regarding the six perspectives~(\faLightbulb[regular]). Afterwards, targets are set and resources were allocated in a prioritization manner regarding the results from previous step~(\faBullseye). 
Then, after gathering additional information from domain experts and stakeholders (groups or individuals affected by planning decisions~\cite{le2016understanding})~(\faBookOpen), initial defined targets are evaluated and adjusted~(\faSliders). After that, detailed planning and design with concrete implementation strategy are developed~(\faCompassDrafting), followed by the plan's realization~(\faChartLine). 

Since real‑world scenarios are dynamic, urban planning process is inherently cyclic. Each iteration refines the approach through learning and adaptation to shifting conditions~\cite{MARASINGHE2024105047}. Planners must balance existing structures with future developments to create sustainable, livable and attractive urban environments~\cite{bibri2017smart}. 

In summary, the ultimate goal remains unchanged to create cities that are not only functional but also sustainable, livable, and attractive. The aim is to design spaces that meet the needs of people today and in the future~\cite{bibri2017smart}. Thus, the aim of this work is to support the urban design process by focusing on the planning phase of new buildings.

\subsection{Style-guidelines for Urban Images} \label{chs:styleguidelines}

This work aims to automatically generate images of buildings with detailed architectural components, such as windows and facade structures, within urban contexts. In this context, to support urban planners during design process, specific style-guidelines for generated images must first be clarified. The following architectural, compositional style-guidelines of urban scenes were determined through interviews with domain experts from \textit{RPTU University Kaiserslautern-Landau, Chair of Architecture}:

\begin{enumerate}[label=\textbf{[C\arabic*]}]
	\item \textit{\textbf{Correctness:}} 
	Buildings and urban environments should be realistic with correct semantic relationships, including detailed architectural components, accurate scale, proportions, and minimal image errors.  This includes detailed depiction of architectural components (e.g. windows), as well as scale and proportional accuracy regarding floor heights, facade structures, window arrangements and material thicknesses. \label{sg:correctness}
	
	\item \textit{\textbf{Completeness:}} 
	The main building should occupy most of the image and clearly connect to its urban context to convey spatial coherence, while one of its four sides must always remain fully visible without clipping, no matter how the image is rotated. \label{sg:completeness}
	
	\item \textbf{\textit{Perspective consistency:}} The image should follow a vanishing-point perspective with all vertical lines kept perfectly straight and parallel to the y-axis, for aesthetic reasons. \label{sg:perspective}
	
	\item \textbf{\textit{Street-View-Level:}} Images should be rendered from street-level in human perspective to capture the overall impression of the environment as experienced by its residents, assuming the scenery is realized as depicted. \label{sg:streetview}
	
	\item \textbf{\textit{Illumination:}} The main building should be optimally lit by optimal light sources, to appear attractive.  This is generally achieved by illuminating the building from the outside, typically through natural light sources like sunlight. \label{sg:illumination}
	
	\item \textbf{Animation:} Humans, animals, and other everyday objects should be included in the urban scene and depicted during their daily activities. \label{sg:vitalization}
\end{enumerate}

Note, that the presented style-guidelines for the image generation of buildings are partially derived from the premise that design inherently encompasses communication and presentation tasks. An example of an optimal presentation image that fulfills all of these guidelines is depicted in Figure~\ref{fig:optimum}.

\section{Theoretical Background}

To understand the proposed MWCFM approach, we first establish the theoretical foundations of FM and WCFM in this section, covering their core principles.

\subsection{Flow Matching}

Inspired by Diffusion models and CFNs, Flow Matching~(FM) constructs a time-dependent velocity field $u_t: \mathbb{R}^d \rightarrow \mathbb{R}^d$ that mediates the transformation of a source distribution $q(x_0) = \mathcal{N}(0, I)$ into a target distribution $q(x_1)$ via integration of an ODE)~\cite{lipman2022flow}. Unlike diffusion-based approaches that rely on iterative noise perturbation and removal, FM directly learns the velocity field along predefined probability paths interpolating between source and target samples. The central advantage is that the optimal velocity field admits an analytical solution along these paths)~\cite{lipman2022flow}. This enables direct regression of a neural network $v_\theta(x_t, t)$ parameterized by learnable parameters $\theta$. Thus, it is trained to minimize the loss

\begin{align}
	\label{eq:fm}
	\mathcal{L}_{\text{FM}}(\theta) &= \mathbb{E}_{t \sim \mathcal{U}[0,1], \, x_t \sim p_t(x)} \left\| v_\theta(x_t, t) - u_t(x_t) \right\|^2 
\end{align}

which measures the mean squared deviation between the velocity field predicted by the network and the analytically determined target velocity field $u_t(x_t)$ of a predefined family of probability densities $p_t$ over all times $t \in [0,1]$. This analytical tractability establishes a deterministic framework in which the learned velocity field guides continuous transformations across probability spaces~\cite{lipman2022flow, calvo2025weighted}. 

Once the model is trained, samples can be generated by numerical integration of the learned ODE using standard solvers~\cite{lipman2022flow}.

The major issue regarding this approach is that the FM loss is computationally intractable in practice, due to the need to integrate over the entire unknown data distribution $p_{\text{data}}$ in order to determine $u_t(x)$ and its corresponding density $p_t(x)$~\cite{lipman2022flow}. However, since we only have access to individual data points (images) from our dataset and not the exact mathematical density function of all existing images, we cannot compute $u_t(x)$ as a target value for the network~\cite{lipman2022flow, calvo2025weighted}.

\subsection{Conditional Flow Matching (CFM)}

Conditional Flow Matching (CFM) decomposes the above described global, intractable problem into numerous simple, tractable sub-problems. Rather than considering the global path of all data, the path on a single, concrete data sample $x_1 \sim p_{\text{data}}$ and on a specific noise sample $x_0 \sim p_0$, is conditioned~\cite{lipman2022flow}. For a single data point pair, the vector field can be computed trivially in closed form. The FM loss from Equation \ref{eq:fm} turn then to the following CFM loss:
\begin{align}
	\label{eq:cfm}
	\mathcal{L}_{\text{CFM}}(\theta) &= \mathbb{E}_{\substack{
			t \sim \mathcal{U}[0,1], \\ 
			x\sim p_{t}(x|x_{1}), \\ 
			x_{1}\sim p(x_{1})
	}} \left[ \left\| v_{\theta}(x_{t}, t) - u_{t}(x|x_{1}) \right\|^2 \right] 
\end{align}

where $u_t(x \mid x_1)$ is refereed as conditional vector field~\cite{lipman2022flow}. Lipman et al.~\cite{lipman2022flow} have proven that the gradients of $\mathcal{L}_{\text{FM}}$ and $\mathcal{L}_{\text{CFM}}$ with respect to the network parameters $\theta$ are exactly identical. Thus, when the model learns to steer toward many random, individual data points $x_1$, it automatically learns, on average the correct global vector field $u_t(x)$~\cite{lipman2022flow}. This process is illustrated in Figure~\ref{fig:fm}.

\subsection{Weighted Conditional Flow Matching (WCFM)}

The performance of neural network training, particularly for FM, depends strongly on the characteristics of the input data. Depending not only on the training objective but also on the complexity of the features to be learned, some samples provide more informative learning signals than others. Calvo-Ordoñez et al.~\cite{calvo2025weighted} address this observation in their \textit{Weighted Conditional Flow Matching~(WCFM)} approach by introducing a weighting function that biases the learned vector field toward regions of interest in the conditional space. This technique extends CFM by applying a non-negative weighting function $w_t(x, x_1)$ to the conditional $ p_t(x\mid x_1) $ on $ t\in[0,1] $ to identify paths that are more important for training. According to Calvo-Ordoñez et al.~\cite{calvo2025weighted} the CFM loss from Equation~\ref{eq:cfm} than turns to~\begin{align}
	\label{eq:wcfm}
	\mathcal{L}_{\text{WCFM}}(\theta) &= \mathbb{E}_{\substack{
			t \sim \mathcal{U}[0,1], \\ 
			x\sim p_{t}(x|x_{1}), \\ 
			x_{1}\sim p(x_{1})
	}} \left[ w_{t}(x,x_{1}) \left\| v_t(x_{t}, t) - u_{t}(x|x_{1}) \right\|^2 \right] 
\end{align}

where $v_\theta(x_t, t)$ and $u_t(x \mid x_1)$ are defined as in CFM, and $w_t(x, x_1)$ is a weighting function that re-weights the contribution of individual samples. Calvo-Ordoñez et al.~\cite{calvo2025weighted} employ a Gibbs kernel as the weighting function:

\begin{equation}
	w(x_0, x_1) = \exp\left(-\frac{c(x_0, x_1)}{\varepsilon}\right),
\end{equation}

where $c$ denotes a cost function and $\varepsilon$ is a regularization parameter controlling the width of the kernel. Intuitively, larger values of~$w(x_0, x_1)$ correspond to sample paths of higher importance for the training objective. For additional theoretical details, we refer the reader to the work of Calvo-Ordoñez et al.~\cite{calvo2025weighted}.
\section{Methodology} \label{ch:image} 

Generating compliant urban building images, which follows the urban style-guidelines from Section~\ref{ch:urban}, requires addressing several challenges regarding model training. Therefore, the aim of this work is to optimize GenAI model training techniques for such image generation tasks. Based on Flow Matching GenAI model type, we propose \textit{Mask-based~Conditional-Weighted~Flow~Matching}~(MWCFM), a expand approach, that incorporates the application-specific urban style-guidelines into FM model training. In this section, we presents our training methodology, as well as the used dataset for urban scene image generation. To our knowledge, this is the first work to employ such an approach.

\subsection{Mask-based Weighted Conditional Flow \\ Matching~(MWCFM)}

We compute the distance on a pixel level within a designated mask~$ m \in M\times N $ and use it to define a weighting factor for the FM loss, thereby extending the WCFM method by Calvo-Ordo\~{n}ez et al.~\cite{calvo2025weighted}. The weighting factor is implemented through the pixel-based local weighted SSIM and local weighted PSNR similarity metrics, with the local weighting induced by the mask. This allows the model to place more emphasis on the relevant region of the image during training, while reducing the influence of irrelevant background areas. In this way, the training objective focuses more strongly on the masked content and can encourage better learning of the target structure. 

Thus, the WCFM loss from Equation~\ref{eq:wcfm} turn to the following MWCFM loss:

\begin{align*}
	\mathcal{L}_{\text{MWCFM}}(\theta) &= \mathbb{E}_{\substack{
			t \sim \mathcal{U}[0,1], \\ 
			x\sim p_{t}(x)
	}} \left[ w_{t}(x_{t}, x_{1}, m^{x_{1}}) \left\| v_\theta(x_{t}, t) - u_{t}(x\mid x_{1}) \right\|^2  \right] 
\end{align*}

with the following weighting function:

\begin{equation*}
	w_{t}(x_{t},x_{1}, m^{x_{1}}) = \lambda_{1}(1-SSIM_{w}(x_{t},x_{1})) + \lambda_{2} \dfrac{1}{PSNR_{w}(x_{t},x_{1})}
\end{equation*}

Here, \(\theta\) denotes the trainable parameters of the vector field model \(v_{\theta}\). The time variable \(t\) is sampled uniformly from \([0,1]\), while \(x\) is drawn from the intermediate distribution \(p_t(x)\). The variable \(x_1\) denotes the target data sample, and \(x_t\) represents the corresponding sample state at time \(t\). The model prediction is given by \(v_{\theta}(x_t, t)\), which is trained to match the target vector field \(u_t(x)\) using MSE loss. The mask \(m^{x_1}\) defines the region of interest associated with \(x_1\) and enters the weighting function \(w_t(x_t, x_1, m^{x_1})\), which modulates each sample’s contribution according to both the mask and the current state. 

Our weighting function is defined based on two complementary similarity measures, \textit{SSIM} and \textit{PSNR}, where \(SSIM_w\) and \(PSNR_w\) denote their weighted counterparts. Although \(SSIM_w\) is formulated for the image pair \((x_t, x)\) as a whole, it is computed from locally evaluated measurements aggregated over the image domain. The parameters \(\lambda_1\) and \(\lambda_2\) regulate the relative contribution of each metric. Since SSIM and PSNR capture distinct aspects of image quality and are defined on different scales, a convex combination of these weights is not appropriate, as it would introduce an undesirable coupling between the terms and impair meaningful metric balancing.

SSIM is employed due to its alignment with human perceptual assessment, as it evaluates luminance, contrast, and structural similarity~\cite{wang2004image}. PSNR, on the other hand, reflects the signal-to-noise relationship and thus provides an indicator of the reconstruction fidelity between the transformed image \(x_t\) and its target \(x\)~\cite{yao2005contrast}.

\subsection{Dataset} \label{chs:dataset}

Choosing a appropriate dataset for machine learning tasks is of great importance, as the input data and the way it is preprocessed strongly influence model training~\cite{gong2023survey}. As a result, the more style-guidelines (see Section~\ref{chs:styleguidelines}) the dataset meets, the more suitable it is for our application task. 

In general, there exist two main categories of datasets: \textit{real datasets} which consists of measured data that come from real sensors (e.g., LiDAR), and \textit{synthetic datasets} which consist of simulated data derived from mathematical models or algorithms regarding a specific task~\cite{jordon2022synthetic}. Currently, neither real open source urban datasets nor synthetic datasets exist that fulfill the proposed requirements~\cite{jordon2022synthetic}. For the presented prototype, we stick to the improvement of optimization of a GenAI model which is able to create images of urban scenes by integrating urban style-guidelines within the training. To generate more realistic images, we decided to use the real dataset Mapillary Metropolis dataset~\cite{mapillary} which provides street level images, measured via LiDAR sensor on top of a driving car. This dataset is divided into the five classes \textit{front cam}, \textit{back cam}, \textit{right cam}, \textit{left cam} and \textit{equirectengular cam}.
Unfortunately, due to camera rotation, strong coverture and lens flare artifacts, some images are partly skewed and defective. Especially, images originating from equirectengular cam cannot be used for training, due to strong curvature. To make the data more feasible regarding the application field, several data cleaning, filtering and correction steps were applied. Another drawback of this dataset is, that the vehicle is visible on images originating from \textit{front cam} or \textit{back cam} classes. 

\subsection{Training}

Training FM models requires deep understanding of machine learning architectures, handling hyperparameters and mathematical foundation in FM theory. In order to generate meaningful images, the model’s architectural complexity must match the complexity of the images. Networks that are too small or incorrectly configured result in heavily noisy images. A similar behavior occurs in hyperparameter handling, especially regarding the learning rate where a notable sensitivity was observed in FM methodologies during research. Suboptimal learning rate configurations frequently resulted in learning failures, manifesting as strongly noisy output across all epochs. For our research, we used the U-Net model from TorchCFM library, developed by Atong et al.~\cite{atong01}, as model architecture.

Training a prototype model using our MWCFM approach, as well as a CFM reference model for validation, using identical architecture and hyperparameters for direct comparison. By integrating urban planning style-guidelines into the loss function, MWCFM targets relevant features more effectively than CFM. To validate our approach, we trained both, an MWCFM prototype and a CFM reference model. Note that, using $ 256 \times 256 $ images, MWCFM required additional memory, due to mask information, but achieved comparable training time per epoch. While the prototype focuses on integrating style-guidelines into training rather than producing publication-ready images, it demonstrates the effectiveness of our methodology.

\subsection{Sampling}

To generate images from the learned distribution, ODE solvers are typically used to integrate the learned ODE~\cite{lipman2024flow}. We use Runge-Kutta as ODE solver for this integration within our prototype model, as well as for the reference model. Both, the MWCFM and the CFM model, used identical sampling methods, architectures, and parameters for fair comparison, with performance evaluated via five application-specific metrics.

\section{Evaluation} \label{ch:eval}

\begin{table*}[!ht]
	\centering
	\begin{tabular}{C{1cm}C{2cm}C{2cm}C{2cm}C{2cm}C{2cm}C{1cm}C{2cm}}
		\textbf{Md.$\backslash$ Ep.} & \textbf{0} & \textbf{5} & \textbf{10} & \textbf{20} & \textbf{30} & \dots & \textbf{50} \vspace{.5cm}\\
		\toprule
		\toprule
		\multirow{-5}{*}{\textbf{MWCFM}} &
		\fbox{\includegraphics[width=0.9\linewidth]{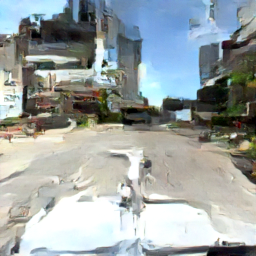}} &
		\fbox{\includegraphics[width=0.9\linewidth]{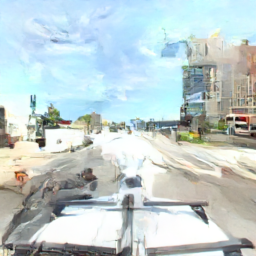}} &
		\fbox{\includegraphics[width=0.9\linewidth]{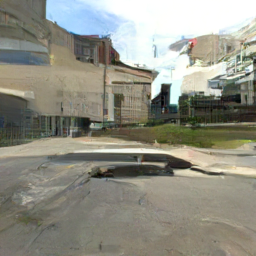}} &
		\fbox{\includegraphics[width=0.9\linewidth]{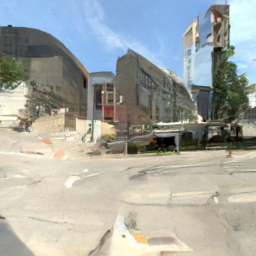}} &
		\fbox{\includegraphics[width=0.9\linewidth]{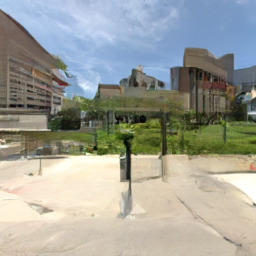}} &
		\multirow{-5}{*}{\dots} &
		\fbox{\includegraphics[width=0.9\linewidth]{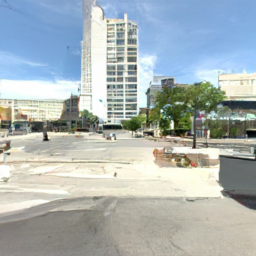}} 
		\\ \midrule
		
		\multirow{-5}{*}{\textbf{CFM}} &
		\fbox{\includegraphics[width=0.9\linewidth]{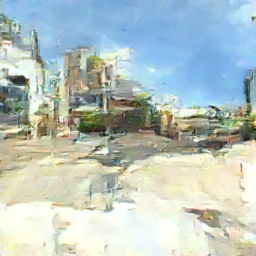}} &
		\fbox{\includegraphics[width=0.9\linewidth]{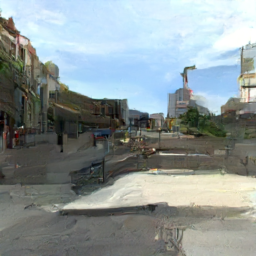}} & 
		\fbox{\includegraphics[width=0.9\linewidth]{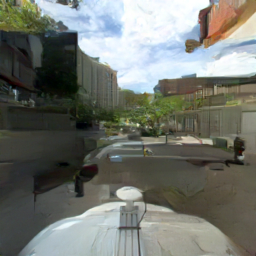}} & 
		\fbox{\includegraphics[width=0.9\linewidth]{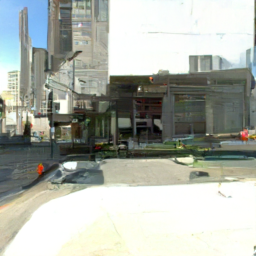}} & 
		\fbox{\includegraphics[width=0.9\linewidth]{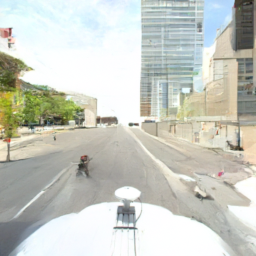}} &
		\multirow{-5}{*}{\dots} &
		\fbox{\includegraphics[width=0.9\linewidth]{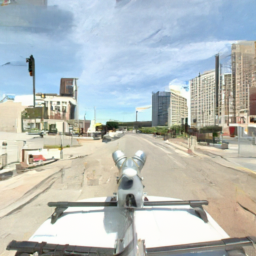}}	
		\\	
	\end{tabular}
	\caption[Generated images from the MWCFM model and CFM reference model]{Generated images from the MWCFM prototype and the CFM reference model of epochs 0, 5, 10, 20, 30 and 50 trained on Mapillary Metropolis dataset~\cite{mapillary}. Learning homogeneous surfaces, like the sky, is fast, as well as reconstructing the vehicle. In images, derived from the MWCFM model, details like windows are recognizable earlier which indicates successful training focus due to the mask-weights.}
	\label{tab:sampling}
\end{table*}

While evaluating the MWCFM model and the CFM reference model, it has to be checked, whether the learned distributions cover the real data distribution, and whether the generated images fulfill the style-guidelines (see Section~\ref{chs:styleguidelines}). Therefore, six scores are calculated, where each score~$ s $ was normalized such that $ s \in [0,100] $, weighted depending on their relevance regarding the application and summed up into a final total score $ s_{total} \in [0,100] $.

\subsection{Evaluation of Model Distribution}

To evaluate if the learned distribution fits the real data distribution the Fr\'{e}chet Inception Distance~(FID) metric was used: 

\begin{enumerate}[label=\textbf{[E\arabic*]}]
	\setcounter{enumi}{-1}
	\item \textbf{FID}: Extracts features from the images and applies Wasserstein distance on it~\cite{de2024reviewing, yu2021frechet} \label{em:fid}
\end{enumerate}   
 
 Here, the ground truth target distribution can be estimated from the real dataset directly. For FM models, the synthetic target distribution is determined by the newly generated samples obtained through numerical integration of the learned ODE. 
 
 Comparing the these two distributions requires that both be estimated from their respective sample sets. Unfortunately, this process inherently introduces several sources of uncertainty. First, the quality of the distribution estimates depends on both, the approximation technique employed, and the number of samples used from each distribution. Even in the idealized scenario where we have access to a perfect, unbiased real dataset and use all available samples to estimate the real distribution, it remains challenging to determine an appropriately sized set of generated samples for suitable approximation of the synthetic target distribution, for reliable comparison.
 
 A second source of uncertainty arises from the sample generation process itself. To generate samples using a FM model, the learned ODE must be solved numerically using ODE solvers such as Runge-Kutta. These solvers introduce their own approximation errors, meaning that even a perfectly trained FM model will produce samples that contain defects stemming from the numerical integration scheme or its configuration.

\subsection{Evaluation of Style-guidelines}

To evaluate the style-guidelines \textbf{\textit{correctness}~\ref{sg:correctness}, \textit{completeness}~\ref{sg:completeness}, \textit{perspective consistency}~\ref{sg:perspective}, \textit{street-view-level}~\ref{sg:streetview}, \textit{illumination}~\ref{sg:illumination}} and \textit{\textbf{animation}}~\ref{sg:vitalization}, we developed five urban metrics for measurement. Unfortunately, due to data defects, generating images that fulfill \ref{sg:completeness}, \ref{sg:illumination} and \ref{sg:vitalization} was not possible. Furthermore, for presentation images, human perception has to be optimized.

\subsubsection{Image Quality Metrics}

First, three basic image quality metrics were developed to get the overall impression:
  
\begin{enumerate}[label=\textbf{[E\arabic*]}]
	\item \textbf{Line Straightness - \ref{sg:correctness}, \ref{sg:perspective}}: \\
	Evaluates geometric correctness and aesthetic quality by extracting line segments via Canny Edge Detection~\cite{canny_edge_CV, opencv_library} and Hough Transform~\cite{hough_line_transform, opencv_library}, then measuring perpendicular distances~\cite{kumar2021shape} between each point and corresponding line, and finally aggregates these into a final score using \textit{Root~Mean~Square~(RMS)}. \label{em:line_straightness}
	
	\item \textbf{Sharpness - \ref{sg:correctness}}: \\
	Quantifies image detail by applying Laplacian filtering to detect regions with rapid intensity changes (e.g. edges) and calculating the variance to measure the deviation of the Laplacian values from their mean. \label{em:sharpness}
	
	\item \textbf{Noise - \ref{sg:correctness}}: \\
	Quantifies detail quality by comparing original pixels with a locally smoothed version using a median filter. After calculating pairwise absolute distances, \textit{median absolute deviation~(MAD)} was applied to get the final score. \label{em:noise}
\end{enumerate}

\subsubsection{Perspective Metrics}

For our application, perspective is very important due to functional as well as aesthetically reasons. To evaluate the perspective related style-guidelines \textit{perspective consistency}~\ref{sg:perspective} and \textit{street-view-level}~\ref{sg:streetview}, we developed the following two metrics.

\begin{enumerate}[label=\textbf{[E\arabic*]}]
	\setcounter{enumi}{3}
	\item \textbf{Vertical Line Metric- \ref{sg:perspective}}: \\
	Checks if the vertical lines within the image are straight and parallel to the y-axis. This is done, by filtering potentially vertical line segments from~\ref{em:line_straightness} and measuring the angle between each of these line segments to the y-axis afterwards. The results are then averaged to get a single score. \label{em:vertical_line}
	
	\item \textbf{Street View Level~(SVL) Metric - ~\ref{sg:perspective}, \ref{sg:streetview}}: \\
	evaluates if the image is in vanishing point, as well as in first-person perspective. This metric consists of two parts:
	
	\begin{enumerate}[left=0pt, label=\arabic*.]
		\item \textbf{Vanishing Point Extraction:}  Intersects line segments from Hough Transform~\cite{hough_line_transform, opencv_library} and clusters intersections via DBSCAN~\cite{dbscanskilearn} to identify vanishing points. 
		
		\item \textbf{SVL Determination:} Measuring vanishing point distances to the horizon line, extracted from filtered horizontal line segments and filtered regarding image position. 
	\end{enumerate}
	\label{em:streetlevel}
\end{enumerate}

\section{Results} \label{ch:results}

For sampling, we used the Runge-Kutta method as ODE solver for both models. Table~\ref{tab:sampling} shows generated image examples at epochs $0$, $5$, $10$, $20$, $30$, and $50$ for the models MWCFM and CFM, trained on Mapillary Metropolis dataset~\cite{mapillary}. Both models quickly learn to reconstruct homogeneous regions such as the sky, although CFM converges slightly faster. Between epochs 10 and 20, the image errors in these areas are already minimal. Both models also perform well on vehicle reconstruction. Interestingly, MWCFM captures building details earlier. So, by epoch 20, windows and other architectural elements are already clearly visible in its outputs, whereas CFM still produces blurred building structures at the same stage.

Both, the MWCFM and the CFM reference model, were evaluated using metrics \textit{\textbf{FID}}~\ref{em:fid}, \textit{\textbf{Line Straightness}}~\ref{em:line_straightness}, \textit{\textbf{Sharpness}}~\ref{em:sharpness}, \textit{\textbf{Noise}}~\ref{em:noise}, \textit{\textbf{Vertical Line}}~\ref{em:vertical_line} and \textit{\textbf{Street View Level}}~\ref{em:streetlevel}. The overall evaluation results for all of this metrics regarding our MWCFM and the CFM reference model can be found in Table~\ref{tab:Eval}. Each metric results was normalized into a single score $ s \in [0,100] $. Here, $0$ indicates poor performance and $100$ represents optimal values.

We generated same amount of synthetic samples for both models. Because FID measures the divergence between distributions, we computed FID score for the MWCFM and the CFM model relative to ground truth. Both models achieved high FID scores, demonstrating strong alignment with ground truth images. MWCFM performs better than CFM in nearly all application-specific metrics, demonstrating its usefulness regarding application-specific training guidance.  Since, the training dataset has inherent limitations~(Section~\ref{chs:dataset}), training performance is limited as well.

\section{Discussion} \label{ch:discussion}

The evaluation results, shown in Table~\ref{tab:Eval}, proof the effectiveness of our approach regarding the presented urban planning presentation image generation task. It can be seen that MWCFM outperforms  standard CFM on \textit{Line Straightness}~\ref{em:line_straightness}, likely due to better handling of complex components. The reason for this could be due to the focus on complex components. However, both models exhibit poor \textit{Sharpness}~\ref{em:sharpness} scores and come with artifacts and blur in generated regions, probably leading this result. On the other hand, the \textit{Noise}~\ref{em:noise} score performs excellent for both. The \textit{Vertical Line}~\ref{em:vertical_line} metric shows high scores for both models. However this should be interpreted in a critical manner, because severely skewed lines fall outside the detection threshold, meaning that they are not detectable as potentially vertical lines within the image. 
\vspace{0.3cm}
\begin{table}[hb]
	\centering
	\begin{tabular}{C{0.5cm}L{2.5cm}C{2cm}C{2cm}}
		&& \textbf{MWCFM} & \textbf{CFM} \\
		\toprule
		\ref{em:fid}&\textbf{FID} & 	97.85 & 98.60 \\ \midrule
		\ref{em:line_straightness}&\textbf{Line Straightness} & 68.83 & 57.91  \\ \midrule
		\ref{em:sharpness}&\textbf{Sharpness} & 24.78 & 22.36  \\ \midrule
		\ref{em:noise}&\textbf{Noise} & 94.63 & 94.74  \\ \midrule
		\ref{em:vertical_line}&\textbf{Vertical Lines} & 91.43 & 90.03  \\ \midrule
		\ref{em:streetlevel}&\textbf{Street View Level} & 25 & 13 
	\end{tabular}
	\caption[Evaluation results for the MWCFM- and the CFM model]{Evaluation results for the MWCFM- and the CFM model. Here, six scores were applied for evaluation. Each score $ s $ was normalized such that $ s \in [0,100] $. It can be seen that both models match the real distribution well, but MWCFM performs better than CFM in nearly all application-specific metrics.}
	\label{tab:Eval}
\end{table}

Unfortunately, the \textit{SVL}~\ref{em:streetlevel} metric reveals more substantial differences. Both models perform poorly here, especially standard CFM. This is probably the case, because vanishing point extraction has to be done in advance to validate the street view perspective which needs successful horizon detection. This fails when artifacts obscure the horizon which is why no meaningful calculation can be done in this case. Additionally, vanishing point extraction from generated images views, originated from \textit{cam front} or \textit{cam back} class, is simpler than from side-camera originated views from classes \textit{cam left} and \textit{cam right}. This is because the camera angle or perspective frequently lacks a clear vanishing point or presents one that is too steep, which can be also the reason why side-camera originated images contain more often artifacts. The unequal class distribution in random generation (vehicle present/absent) further affects this score. Nonetheless, the \textit{SVL}~\ref{em:streetlevel} score of the MWCFM model nearly doubled compared to the score of the CFM model, suggesting fewer disruptive artifacts and higher success rate.

Overall, MWCFM and CFM achieve similar composite scores, with MWCFM performing marginally better. Important caveats apply to \textit{FID}~\ref{em:fid}, because it depends on the numerical approximation algorithm, sample quality and feature-extraction methods, involving uncertainties. Additionally, numerical approximation errors when integrating the learned ODE and inappropriate sampling mechanism selection can degrade synthetic image generation, even assuming a perfect model. In case of FM, this means that in worst case, image generation can result in completely noisy images (e.g., white noise), regardless from quality of the trained model. Debugging is not trivial in this context.

\section{Conclusion} \label{ch:con}

The potential of \textit{Generative Artificial Intelligence}~(GenAI) models has expanded across diverse domains, including urban planning. This study demonstrates that GenAI can be tailored to produce architectural presentation images for urban planning by embedding style constraints, derived from application specific style-guidelines, directly into the training loss. Using Conditional Flow Matching~(CFM) as the underlying GenAI model type, we introduced \textit{Mask-based Weighted Conditional Flow Matching}~(MWCFM), which extends standard FM with mask‑weighted losses that concentrate learning on complex, style‑critical regions.

Comparative experiments with an otherwise identical standard CFM baseline show that MWCFM yields modest gains in both FID and adherence to urban style-guidelines. The improvements are limited by sub‑optimal dataset that does not fully satisfy the prescribed style rules and small image sizes. Thus, the models produce acceptable but not yet production‑ready images. Furthermore, evaluation of such presentation images is still challenging.

Future work should focus on (1) assembling or curating a high‑quality, style compliant dataset, (2) training at higher resolutions to meet architectural presentation standards, and (3) developing an interactive control interface that lets architects and planners steer the generation process toward their specific design intents.

\section*{Acknowledgment}

During the preparation of this work the authors used GPT-5 for language polishing, and Gemini 3 to assist in drafting \textit{matplotlib} code for Figure~\ref{fig:fm}. After using these tools, the authors reviewed and edited the content as needed and take(s) full responsibility for the content of the published article.


\bibliographystyle{abbrv-doi}
\bibliography{references}

\end{document}